\documentclass[a4paper]{jpconf}
\usepackage{graphicx}
\begin{document}
\title{Deep Learning for Fatigue Estimation on the Basis of Multimodal Human-Machine Interactions}

\author{Yu.Gordienko*, S.Stirenko, Yu.Kochura, O.Alienin, M.Novotarskiy, N.Gordienko}

\address{National Technical University of Ukraine "Igor Sikorsky Kyiv Polytechnic Institute", 37 Peremohy Avenue, Kyiv 03056 Ukraine}

\ead{*yuri.gordienko@gmail.com}

\author{A.Rojbi}

\address{CHArt Laboratory (Human and Artificial Cognitions), University of Paris 8, 2 Rue de la Liberté, 93526 Saint-Denis, France}


\begin{abstract}
The new method is proposed to monitor the level of current physical load and accumulated fatigue by several objective and subjective characteristics. It was applied to the dataset targeted to estimate the physical load and fatigue by  several statistical and machine learning methods. The data from peripheral sensors (accelerometer, GPS, gyroscope, magnetometer) and brain-computing interface (electroencephalography) were collected, integrated, and analyzed by several statistical and machine learning methods (moment analysis, cluster analysis, principal component analysis, etc.). The hypothesis 1 was presented and proved that physical activity can be classified not only by objective parameters, but by subjective parameters also. The hypothesis 2 (experienced physical load and subsequent restoration as fatigue level can be estimated quantitatively and distinctive patterns can be recognized) was presented and some ways to prove it were demonstrated. Several ``physical load'' and ``fatigue'' metrics were proposed. The results presented allow to extend application of the machine learning methods for characterization of complex human activity patterns (for example, to estimate their actual physical load and fatigue, and give cautions and advice).
\end{abstract}

\section{Introduction}
Physiological signals can be categorized into cerebral (electroencephalography, functional magnetic resonance imaging, etc.) and peripheral (heart rate, biological activity, temperature, etc.) ones. Due to development of wearable electronics and Internet of Things, these complex physico-chemical signals can be recorded and transmitted from various limbs, and then they can be statistically analyzed as an integral set of multimodal human-machine interactions. Quantifying the above mentioned mental and physical efforts is non-trivial task. The main aim of this paper is to present the new approach to monitor the level of current physical load and accumulated fatigue by several objective and subjective characteristics. The section \emph{2.Background} gives the very brief outline of the state of the art. The section \emph{3.Experimental} contains the short characterization of experimental part related with main terms, parameters and metrics. The section \emph{4.Results} reports about the results obtained and processed by some statistical and machine learning methods. The section \emph{5.Discussion} is dedicated to discussion of the results obtained and lessons learned.

\section{Background}
The problem of estimation of the actual physical load and fatigue appeared from older times~\cite{craighead2004concise} and nowadays it is evolved to the concrete challenges~\cite{Chernbumroong2013Elderly,banaee2013data,lopez-nava2016wearable}, especially in the context of human-machine and machine-human interactions~\cite{vinciarelli2015open}. In older times the technical side of the problem was related with limitations of data monitoring, collecting, processing, and representing by the available tools. But now the rapid development of information and communication technologies allow us to widen the range of the sensors and actuators, which are already become de facto standard devices in the ordinary gadgets~\cite{meng2015review}. For example, sound, light, acceleration, proximity, gesture, haptic, heart rate, and touch sensors are present in the majority of the modern electronic gadgets like mobiles, tablets, etc. The fast increase of available sensors in the recent decades determined the need for many more tacit and informative means for representation of various characteristics of human behavior patterns, especially related with physical activity and subsequent physical fatigue~\cite{lemoyne2010implementation}. It is especially important for elderly care applications~\cite{gordienko2017augmented} on the basis of the newly available information and communication technologies with multimodal interaction through human-computer interfaces like wearable computing, augmented reality, brain-computer interfaces, etc~\cite{stirenko2017user}. Recently several approaches of fatigue estimation were proposed on the basis of multimodal human-machine interaction and machine learning methods~\cite{LeMoyne2015ml}. The valuable output can be obtained by usage of machine learning and, especially deep learning techniques, which are recently used for analysis of human physical activity~\cite{lemoyne2018roleML}. Below, the approaches and metrics for their quantitative estimation are proposed and discussed. 

\section{Experimental}
The proposed method consists in monitoring the whole spectrum of human movements, which can be estimated by Tri-Axial Accelerometer (TAA) and heartbeat/heartrate (HB/HR) monitor in the modern  smartphones, smartwatches, fitness-trackers, or other fitness-related gadgets (like FitBit, Armour39, etc.) with an optional synergy with internal or external sensors in the connected smartphone and data on ambient conditions in the smartphone. The main principle is the paradigm shift: to go from ``raw output data'' (used in many modern accelerometry based activity monitors) to the ``rich post-processed (and, optionally, ambient-tuned) data'' obtained after smart post-processing and statistical (moment/cluster/bootstrapping) analyses with much more quantitative parameters. 

\begin{figure}[!h]
\centering
        \includegraphics[height=7cm]{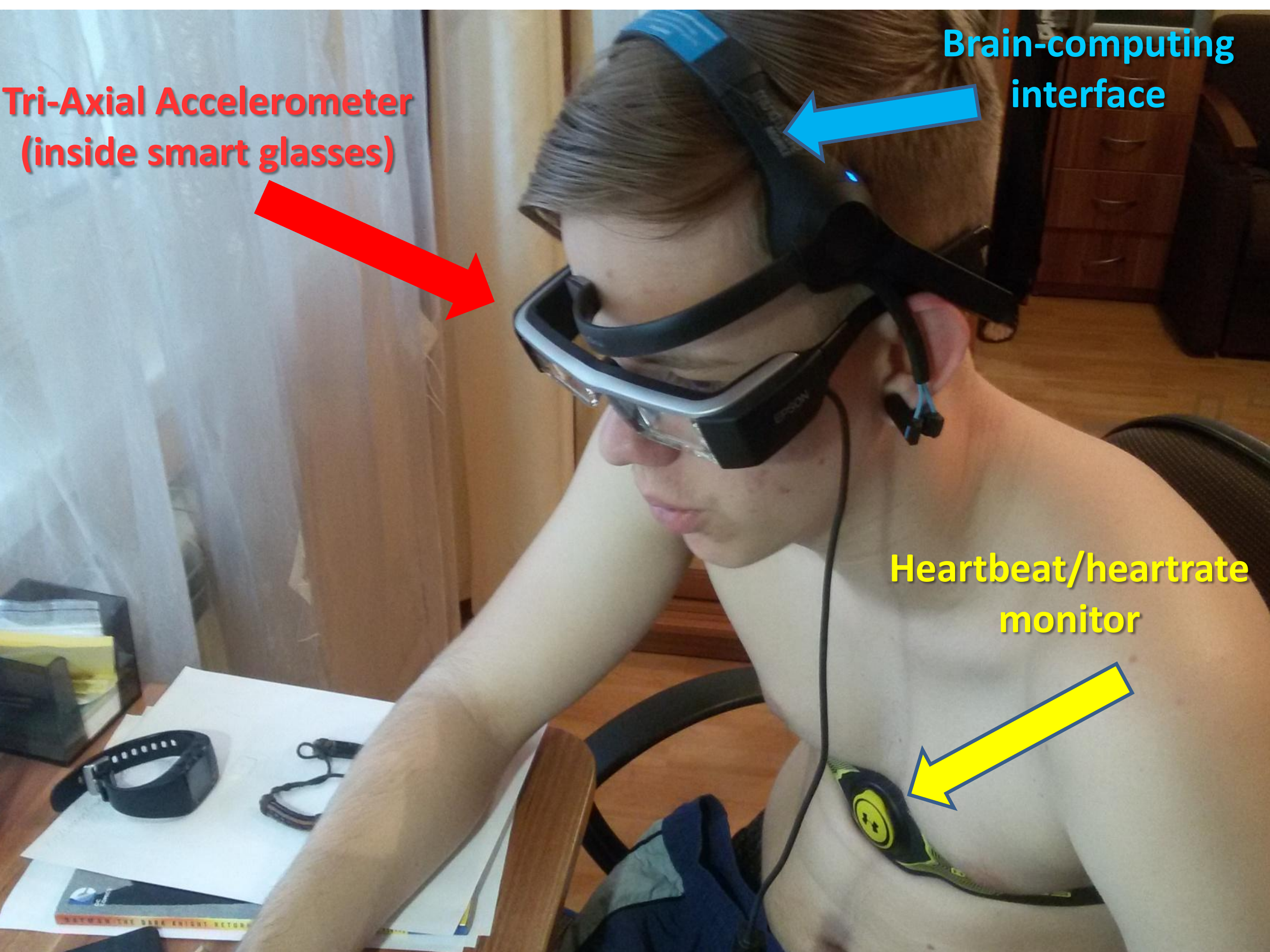} a) \includegraphics[height=7cm]{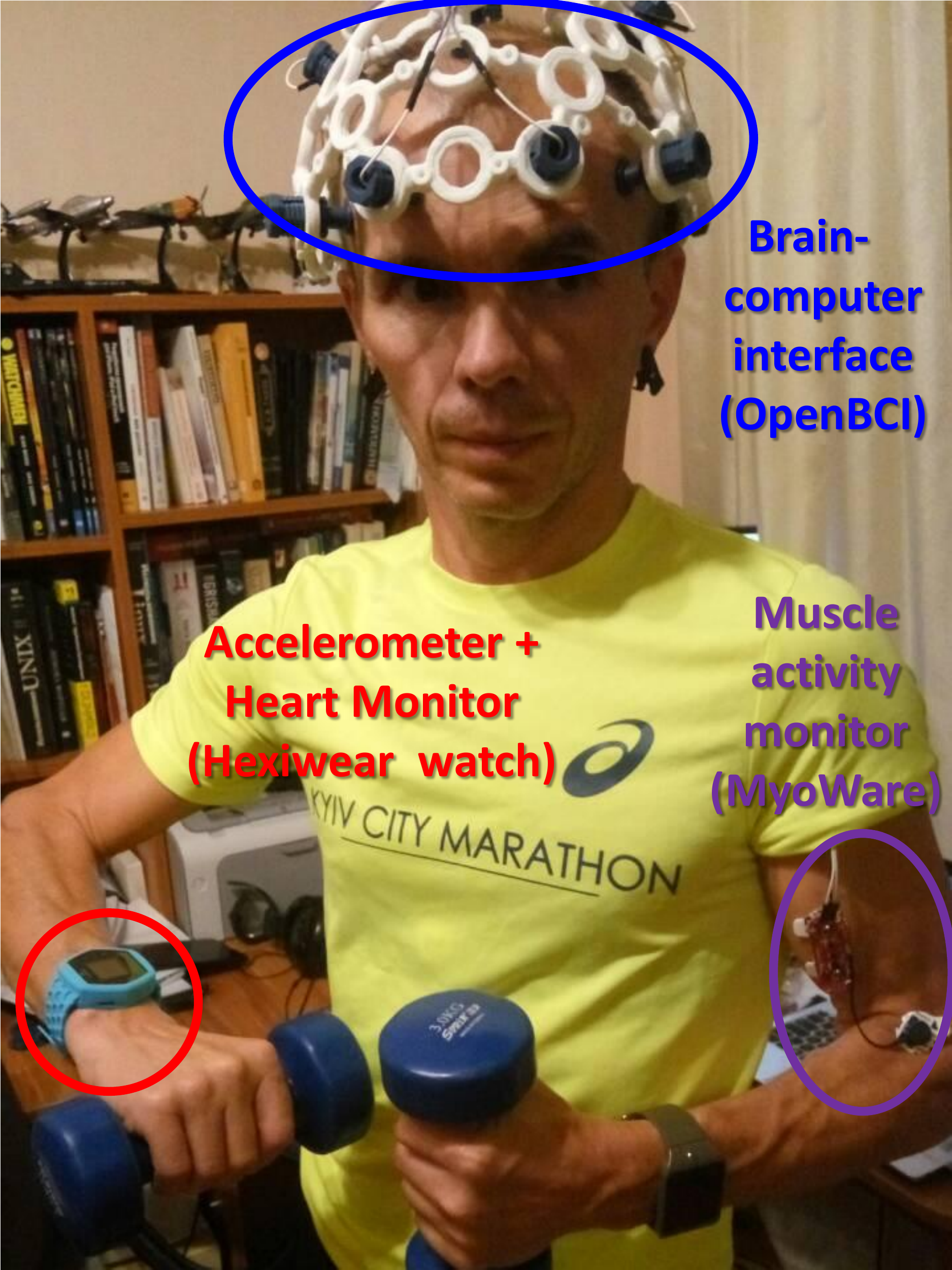} b)
        \caption{Experimental setup to measure mental and physical efforts: (a) basic configuration, and  (b) extended experimental setup to measure brain (brain-computer interface by OpenBCI~\cite{OpenBCI}), muscle (electromyography (EMG) sensors by MyoWare~\cite{Myoware}), and heart activities (heart monitor by Hexiwear~\cite{Hexiwear}) to measure mental and physical efforts(this option is reserved for the future research)~\cite{stirenko2017user}.}
        \label{Fig02_exp_setup}
\end{figure}

The typical usage scenarios for investigation and tests were as follows: a) heart rate and heartbeat (HB/HR) activity; b) acceleration activity for different exercises; c) cumulative information from several sensors, including sensors in smart glasses and brain-computer interface (BCI) on a head (Fig.~\ref{Fig02_exp_setup}b) (not covered in detail here ~\cite{stirenko2017user}).

The measurements of the general activity patterns for the human body (as a whole) and human limbs (separately) were performed on the basis of TAA in smartphone and HB/HR monitor in Armour39 by Under Armour heart rate monitor. The attachment points were used in the same way like it was explained above: a) on the arm (during writing, web-surfing, etc); b) on the limbs during walking and running; c) near the palm; d) on the head (to estimate static fatigue on neck bone). The cheap standard elastic wristbands were used for these tests.

The following experimental setups~(Fig.~\ref{Fig02_exp_setup}a) were used where the person under investigation performed several mental and physical actions and several sensoric data channels were used to measure his/her response. The most promicing feedback as to the mental activity and efforts can be obtained by the advanced experimental setup with the more specific and accurate devices on the basis of multichannel brain-computer interface like OpenBCI, which is an open source brain-computer interface platform, created by Joel Murphy and Conor Russomanno~\cite{OpenBCI} (see Fig.~\ref{Fig02_exp_setup}b). In the similar way the more specific information now is gathered by the locally situated muscle sensors (like Myoware~\cite{Myoware}) and heart rate monitors (breast heart monitor and wrist heart monitor like Hexiwear~\cite{Hexiwear}) (Fig.~\ref{Fig02_exp_setup}b).

\begin{figure}[!h]
\centering
        \includegraphics[height=10cm]{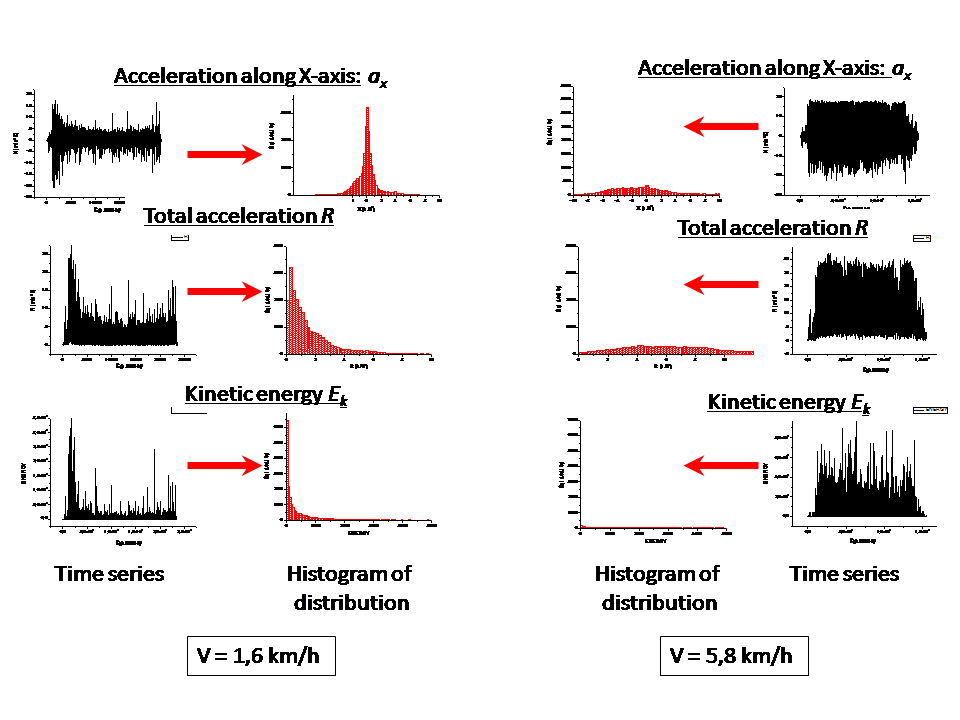}
        \caption{The conceptual idea (the paradigm shift) of the statistical analysis behind the proposed technology: to go from ``raw output data'' (left and right columns) (used in many modern accelerometry based activity monitors) to the ``rich post-processed (and, optionally, ambient-tuned) data'' (central column) obtained after smart post-processing and statistical analysis (with much more quantitative parameters).}
        \label{Fig03_concept}
\end{figure}

The post-processing and statistical analysis of the raw data given by various TAA and HB/HR monitors was performed for various physical activities. Several statistical methods (moment, bootstrapping and cluster analyses), which were successfully used before for other mathematical~\cite{gordienko2014change} and physical ~\cite{gordienko2012generalized,gordienko2011molecular} applications, were applied for the data supplied by TAA and HB/HR monitor. The conceptual idea of the statistical analysis behind the proposed approach is described in the following scheme on Fig.~\ref{Fig03_concept}. Its essence is the paradigm shift: to go from “raw output data” (left and right columns in Fig.~\ref{Fig03_concept}) (used in many modern accelerometry based activity monitors) to the “rich post-processed (and, optionally, ambient-tuned) data” (central column) obtained after smart post-processing and statistical analysis (with much more quantitative parameters).

\section{Results}
\subsection{Accelerometry data channel}
\begin{figure}[!h]
\centering
        \includegraphics[height=6cm]{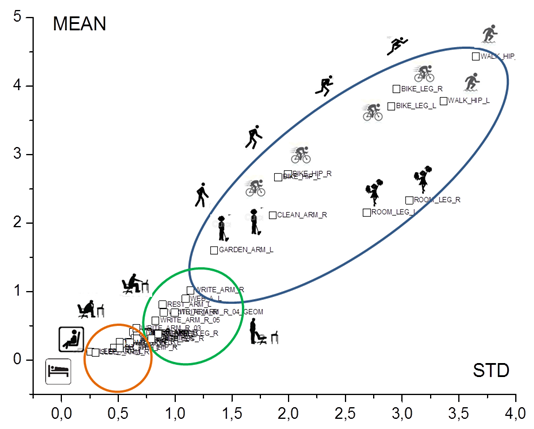}a) \includegraphics[height=6cm]{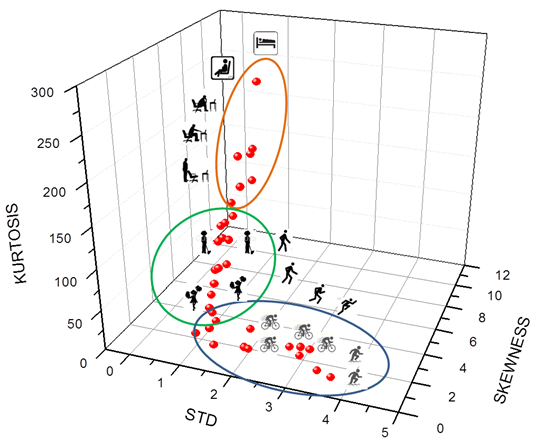}b)
        \caption{Conceptual idea as to multiparametric moment analysis: activities can be classified in more details, i.e. divided into groups (colored ellipses) with the similar values of the acceleration distribution parameters: the active (sports, housework, walking) (blue ellipse), moderate (writing, sitting) (green ellipse) and passive (web surfing, reading, sleeping) (brown ellipse) behavior.}
        \label{Fig04a_ACTIVITY}
\end{figure}

The time series of accelerations measured during movements of human body were considered as statistical samplings. Then the distribution of acceleration values in these samplings were analyzed by calculation of mean, standard deviation, skewness, and kurtosis. Below the experimental data on various human activities (symbolic icons note the type of activity) are shown for: 2 parameters “mean-standard deviation” (Fig.~\ref{Fig04a_ACTIVITY}a) and 3 parameters “kurtosis-skewness-standard deviation (std)” (Fig.~\ref{Fig04a_ACTIVITY}b), which could be extended to the more number of parameters. Due to differences in statistical parameters of these distributions, some activities can be classified, i.e. divided into groups (colored ellipses) with the similar values of the acceleration distribution parameters: the active (sports, housework, walking) (blue ellipse), moderate (writing, sitting) (green ellipse) and passive (web surfing, reading, sleeping) (brown ellipse) behavior (Fig.~\ref{Fig04a_ACTIVITY}).

As one can see, activities can be classified in more details, i.e. divided into groups with the similar values of the acceleration distribution parameters (mean, standard deviation, kurtosis, skewness, etc.) (Fig.~\ref{Fig04a_ACTIVITY}). The different types of activity are denoted also by the symbolic self-explanatory icons.

\subsection{Heart rate/Heart beat data channel}
The similar approach was applied for estimation of the workload during exercises and its influence on heart. The crucial aspects of this approach are as follows: 
\begin{enumerate}
    \item the absolute values of heart rates (heartbeats) for the same workload are volatile and sensitive to the person (age, gender, physical maturity, etc.) and its current state (mood, accumulated fatigue, previous activity, etc.) -- what should be done: in contrary, their distributions should be used instead of some limited values;
    \item the heart rate values are actually integer values with 2-3 significant digits and not adequately characterize the volatile nature of heart activity (because the heart rate is actually the reverse value of the heartbeat multiplied by 60 seconds and rounded to integer value) -- what should be done: in contrary, heartbeats in milliseconds should be used, because they contain 3-4 significant digits and their usage gives 10 times higher precision and much more informative.
\end{enumerate}

\begin{figure}[!h]
\centering
        \includegraphics[height=7cm]{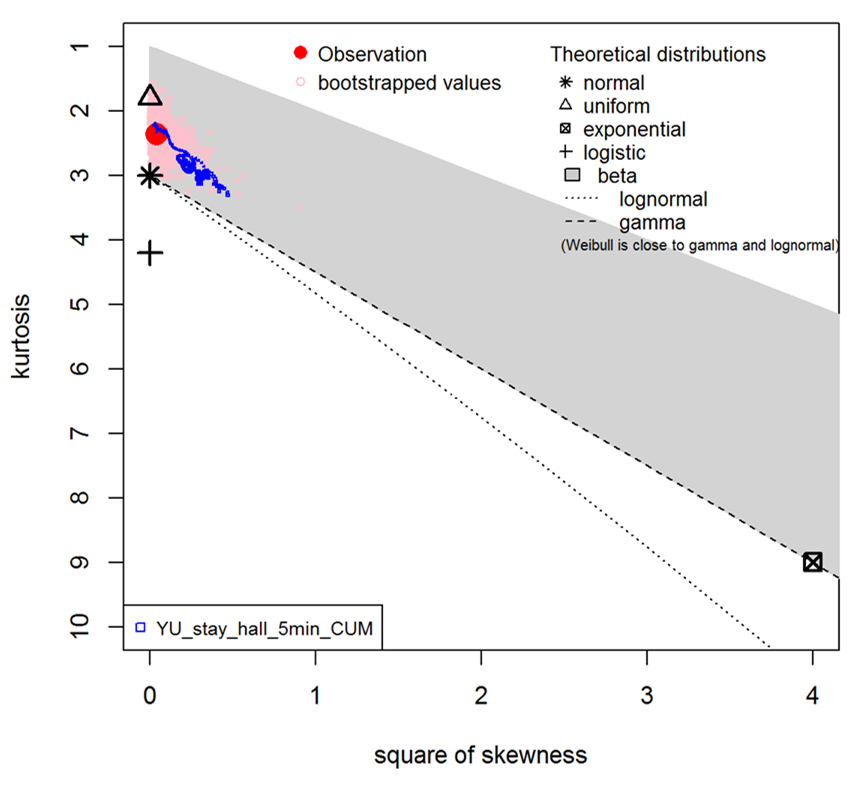} a) 
        \includegraphics[height=7cm]{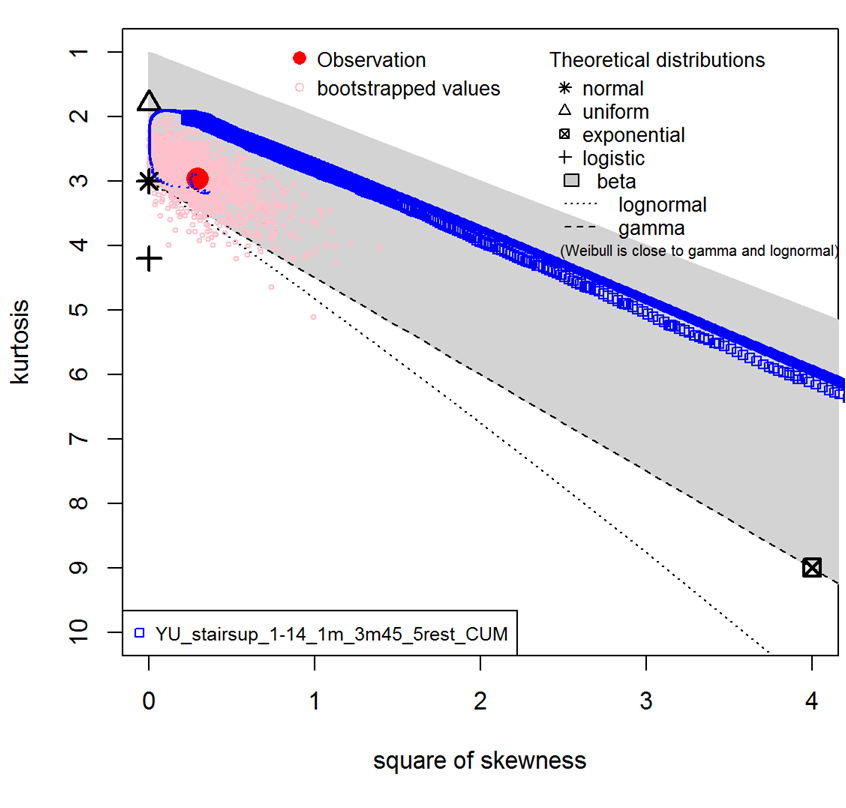} b)
        \caption{Moments diagrams of the heartbeat distribution vs. exercise time for rest staying during 5 min (a), and walking upstairs (b) for the well-trained person (YU, male, 47 years). The exercise was like: 1 min of rest + 3.45 min of walking upstairs (13 floors) + 5 min of rest. Legend: Size of blue symbol (current moments of heartbeat distribution) increases with time of exercise. Red color point -- initial position (near normal distribution). Rose color cloud of points -- results of bootstrapping analysis.}
        \label{Fig05_CF_rest_staircase}
\end{figure}

This method allows us to determine the level of workload and recovery after other types of exercise (for example, walking upstairs up to 13th floor, here) from the moments diagrams of the heartbeat distribution vs. exercise time for various workloads (Fig.~\ref{Fig05_CF_rest_staircase}a) and metrics plots (Fig.~\ref{Fig05_CF_rest_staircase}b). The exercise was like: 1 min of rest + 3.45 min of walking upstairs (13 floors) + 5 min of rest. Again, the slopes of metric increase and decrease can be used to characterize the accommodation and recovery levels during these exercises.

\begin{figure}[!h]
\centering
        \includegraphics[height=6cm]{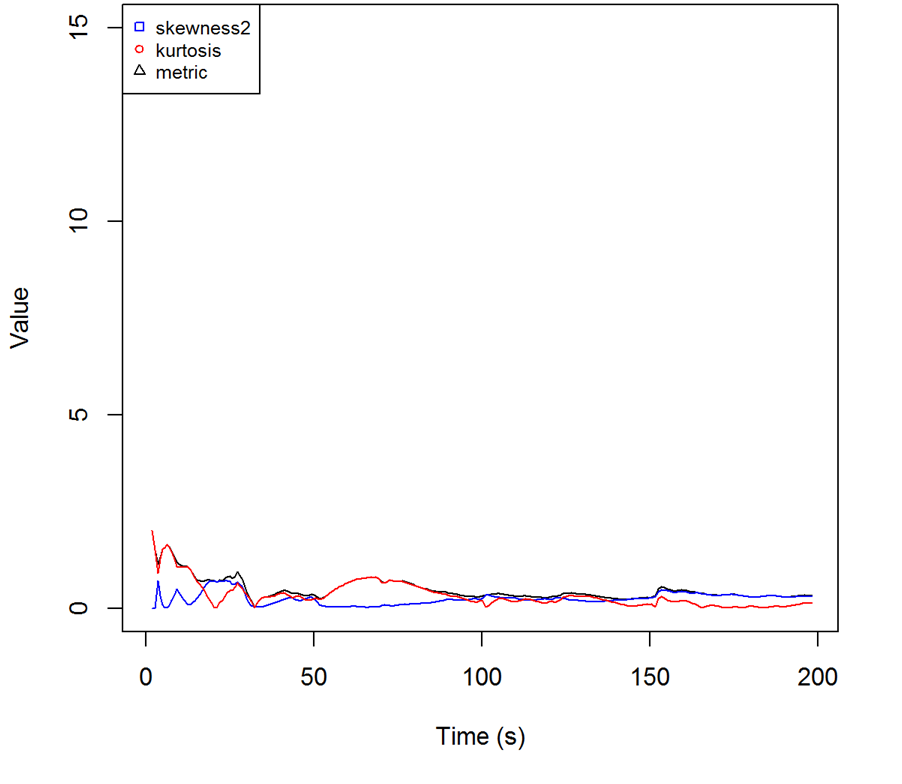} a) 
        \includegraphics[height=6cm]{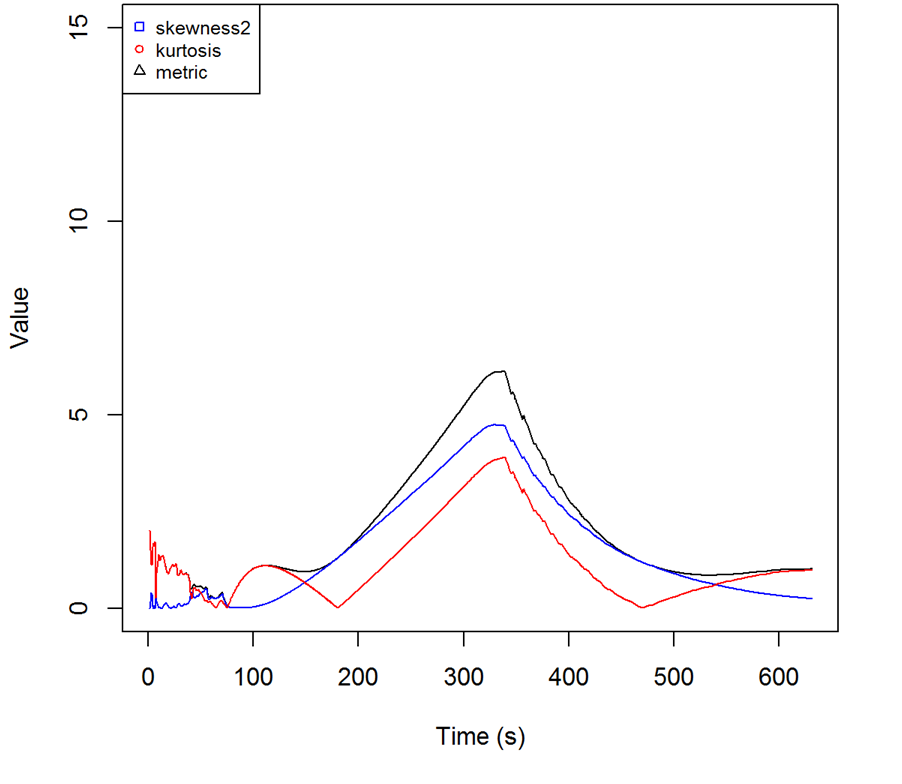} b)
        \caption{Monitoring the adaptation of person to loads by metrics for various typed of activities: (a)~rest state, (b)~walking up stairs. Legend: black line -- Metric1, red line -- kurtosis, blue line -- square of skewness.}
        \label{Fig06_METRIC_rest_staircase}
\end{figure}

The following metrics were used to characterize the accommodation and recovery levels during these exercises: the distance from the normal distribution on the moments diagram (Metric1) and the distance from the uniform distribution on the moments diagram (Metric2). The plots in~Fig.~\ref{Fig06_METRIC_rest_staircase} show evolution of the two metrics of the heartbeat distributions vs. exercise time for rest staying during 5 min (Fig.~\ref{Fig06_METRIC_rest_staircase}a) and exercise (Fig.~\ref{Fig06_METRIC_rest_staircase}b) for the well-trained person (male, 47 years). The exercise was like: 1 min of rest + 3.45 min of walking upstairs (14 floors) + 5 min of rest. 

\begin{figure}[!h]
\centering
        \includegraphics[height=6cm]{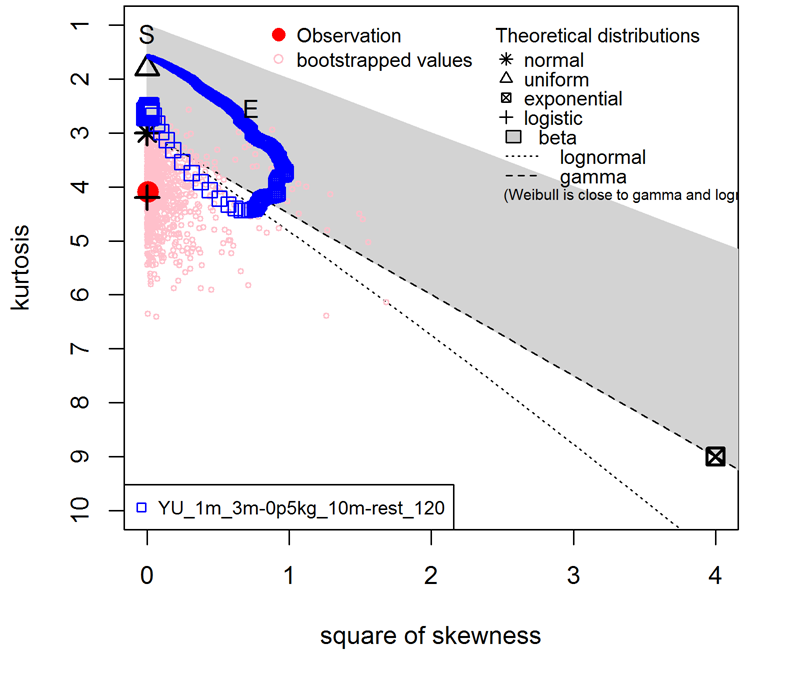} a)
        \includegraphics[height=6cm]{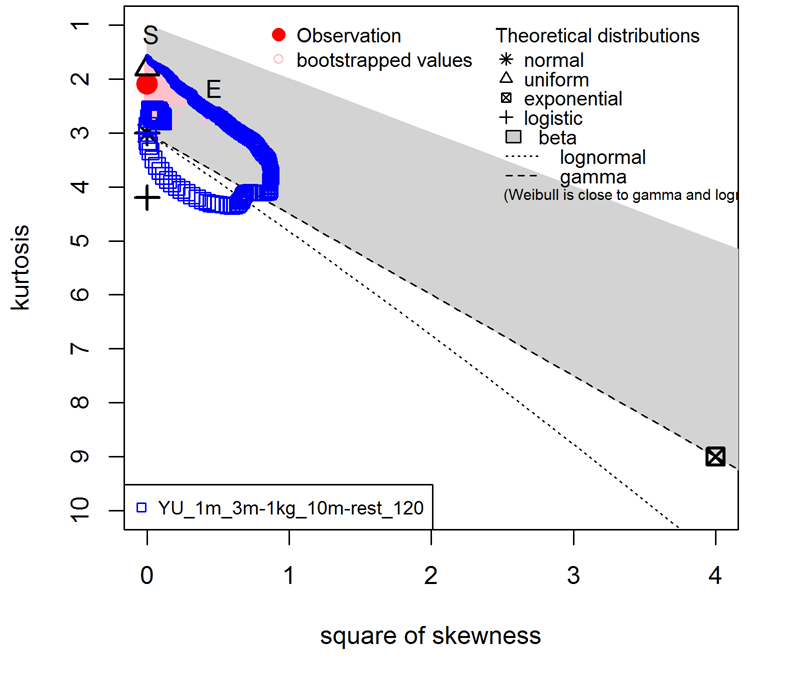} b)
        \includegraphics[height=6cm]{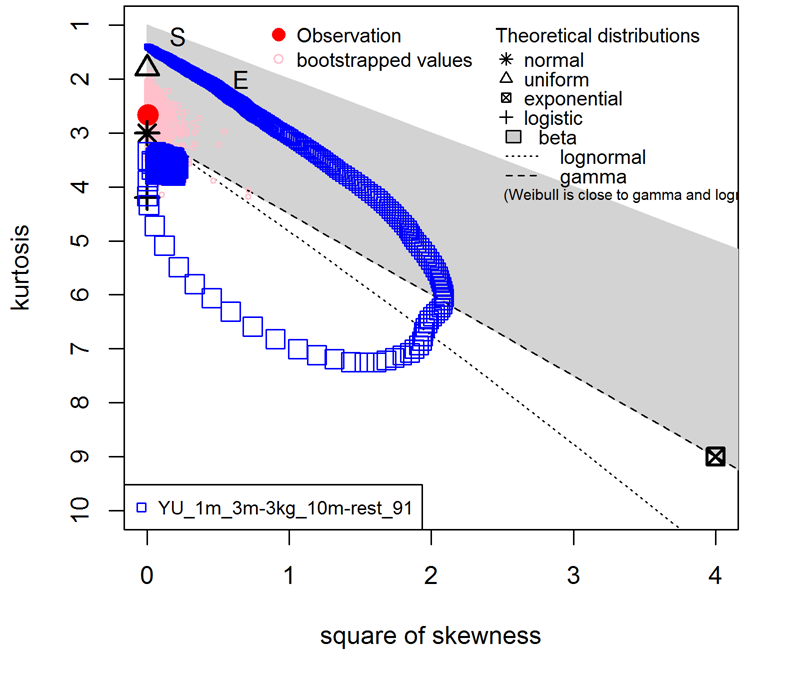} c)
        \caption{Heartbeats vs. exercise time for dumbbell curl for bicep: (a)~lower load (0.5 kg), (b)~medium load (1.0 kg), (c)~higher load (3.0 kg). Exercises: after 1 min of rest, dumbbell curl for biceps up to fatigue, 10 min of rest (recovery phase). Legend: Size of blue symbol (current moments of heartbeat distribution) increases with time of exercise. Red color point -- initial position (near normal distribution). Rose color cloud of points -- results of bootstrapping analysis. S -- start of exercise, and E -- end of exercise.}
        \label{Fig07_CF_dumbbell}
\end{figure}

This method also allows us to observe influence of workload and recovery after exercise from the moments diagrams of the heartbeat distribution functions vs. exercise time for various workloads~Fig.~\ref{Fig07_CF_dumbbell}: a) 0.5 kg (very low load), b) 1 kg (comfort load), c) 3 kg (high load). Each point denotes the sampling containing several hundreds neighboring measurements obtained by sliding window method. During exercise (from the start point S to the end point E in ~Fig.~\ref{Fig07_CF_dumbbell}) the sampled distributions start from the position of the uniform distribution and with load go through the zone of beta-distributions. During relaxation (after the end point E in ~Fig.~\ref{Fig07_CF_dumbbell}) the sampled distributions go through the zone of Weibull distributions and tend to the rest state near the location of the normal distribution. The trajectory for the higher workload (the higher weight of dumbbell) has tendency to the higher curvature in the region of Weibull distribution, which is usually used for description of the numerous critical processes from the point of view of weak-link interpretation~\cite{weibull1951statistical,gordienko2014change,gordienko2011molecular}.

\subsection{Correlation Analysis}
\begin{figure}
\centering
        \includegraphics[width=4.75cm]{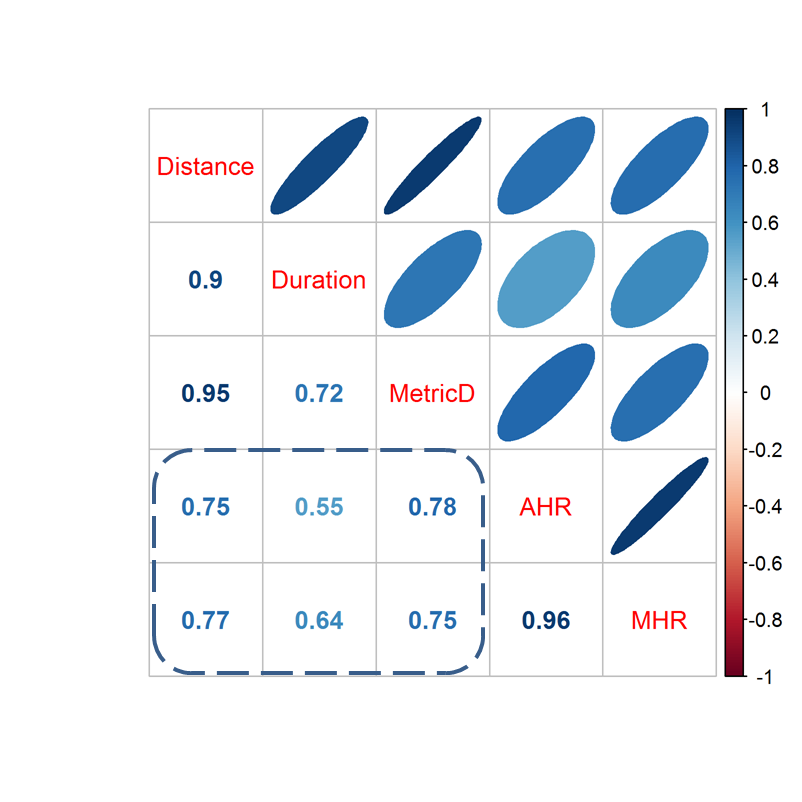}a) \includegraphics[width=4.75cm]{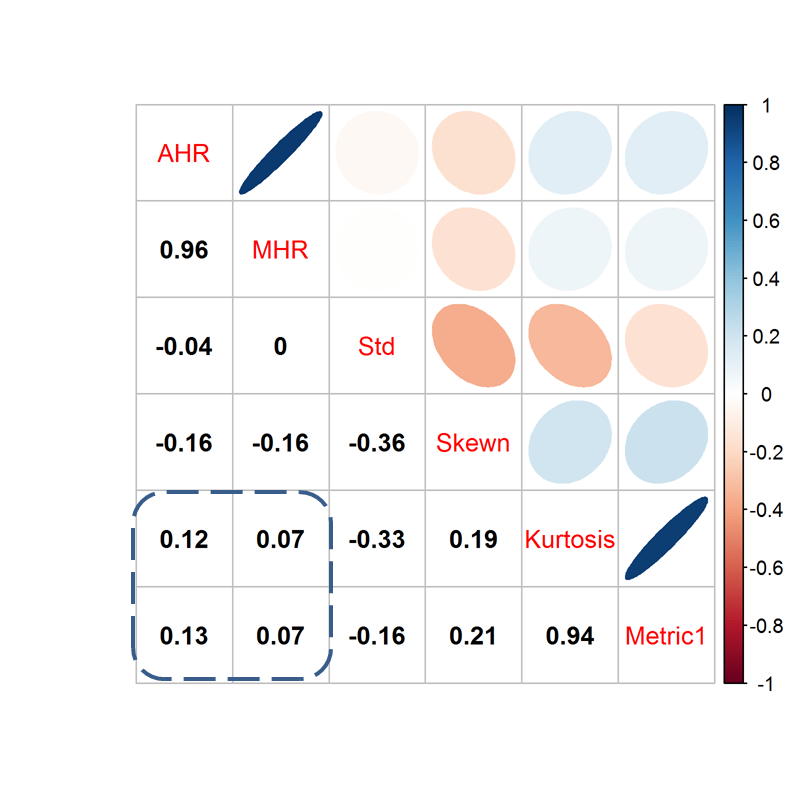}b) \includegraphics[width=4.75cm]{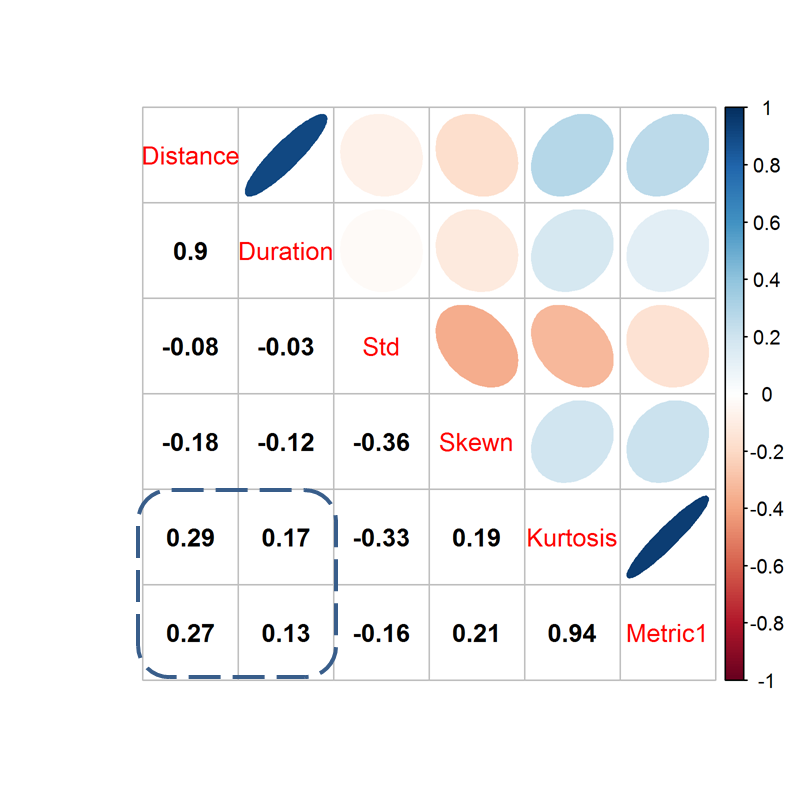}c)
        \caption{Correlation results for the data obtained by multimodal channels like acceleration and 
        heart activity.}
        \label{Fig08_correlation}
\end{figure}

To investigate the potential dependencies (Fig.\ref{Fig08_correlation}) between the observed data and parameters of the physical activities, the correlation analysis was performed for the several parameters of the physical activities of various types (walking, running, skiing):
\begin{itemize}
    \item objective: distance, duration, velocity, pace, and the combined metric based on the previous parameters (for example, MetricD = pace*pace);
    \item subjective: average heart rate, maximal heart rate, statistical parameters of accelertion distributions (mean, std, skewness, kurtosis), and the combined metric based on the previous parameters (for example, Metric1 and Metric 2 that were mentioned above).
\end{itemize}

The well-known correlation between the objective parameters (distance and duration of training) and some heart-related parameters, actually average heart rate AHR) and maximal heart rate (MHR), were observed (Fig.\ref{Fig08_correlation}a). To find the correlation between the heart-related parameters (AHR, MHR) and fatigue after exercise the muscle tremors were measured by accelerometry. Despite the low correlation (Fig.\ref{Fig08_correlation}b) between the heart-related parameters (during exercise) and parameters of acceleration distributions (after exercise), the better correlation (Fig.\ref{Fig08_correlation}b) was observed between the objective parameters (distance and duration of training) and parameters of acceleration distributions (after exercise). These results are not persuasive for any reliable conclusions, but were incentive for the continuation of investigations of links between objective parameters of load and subjective health parameters.

\subsection{Machine Learning}
\begin{figure}[!h]
\centering
        \includegraphics[width=5.25cm]{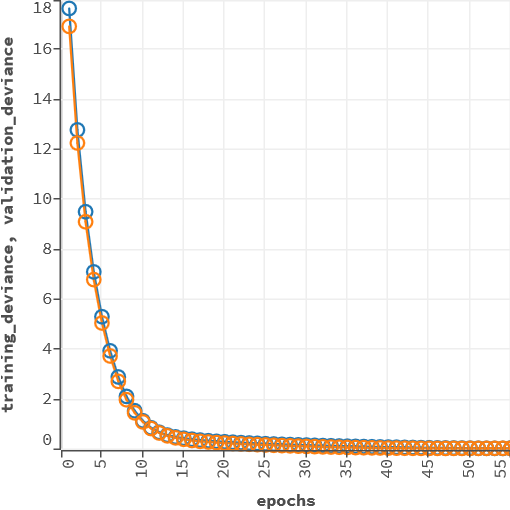} a) \includegraphics[width=5.25cm]{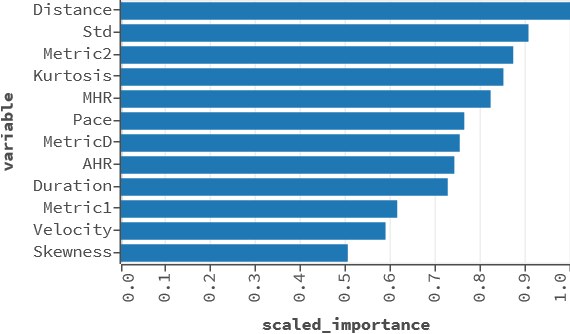} b)
        \includegraphics[width=5.25cm]{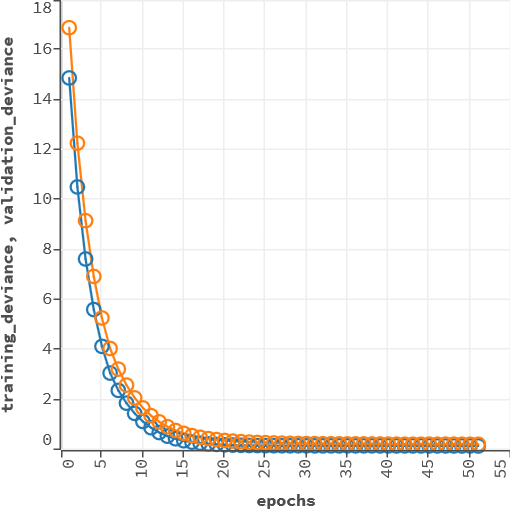} c) \includegraphics[width=5.25cm]{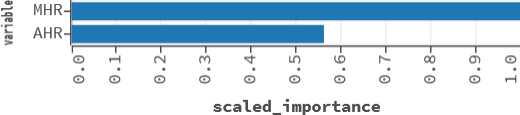} d)
        \includegraphics[width=5.25cm]{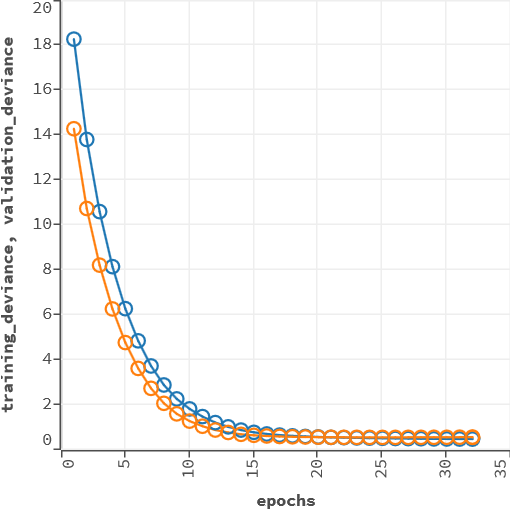} e) \includegraphics[width=5.25cm]{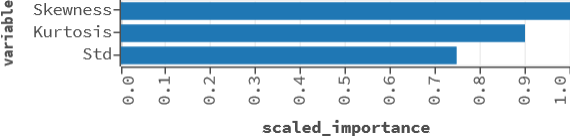} f)
        \caption{The training (blue) and validation (yellow) rates with epochs of machine learning (the left plots), and the key parameters and their relative influence on training the neural network are shown on (the right plots).}
        \label{Fig09a_FATIGUE}
\end{figure}

Recently, various machine learning approaches were applied to characterize the human behavior patterns and perform health and fatigue diagnostics~\cite{LeMoyne2015ml,lemoyne2018roleML}. To investigate the experimental data on human activity several open-source machine learning networks were analyzed, tested~\cite{kochura2017comparative,kochura2017comparativeperformance,kochura2017ysf}, and then applied for linear regression model analysis and deep learning analysis.  The aim was to train the deep learning network on the set of the physical exercise data and predict the type of activity (walking, running, skiing) from the set of the aforementioned objective and subjective parameters.  
The training results obtained by application of various machine learning methods, actually, by deep learning neural networks~\cite{kochura2017ysf}, for the data obtained by multimodal channels (acceleration and heart activity) are shown in Fig.~\ref{Fig09a_FATIGUE}. The training (blue) and validation (yellow) rates with epochs of machine learning are shown on the left plots (Fig.~\ref{Fig09a_FATIGUE}a,c,e), and the key parameters and their relative influence on training the neural network are shown on the right plots (Fig.~\ref{Fig09a_FATIGUE}b,d,f).

The precision of validation and prediction was estimated and characterized by Mean Residual Deviance (MRD) and mean absolute error (MAE) for various combination of objective and subjective parameters (Table~\ref{Table_errors}).

\begin{table}
    \caption{\label{Table_errors}Mean Residual Deviance (MRD) and mean absolute error (MAE) for training, validation, and prediction stages of machine learning shown in Fig.~\ref{Fig09a_FATIGUE}.}
    \begin{center}
        \begin{tabular}{lllll}
            \br
            Parameters &\multicolumn{2}{c}{Validation} & \multicolumn{2}{c}{Prediction}\\\cline{2-3}\cline{4-5}
               & MAE & MRD & MAE & MRD\\
            \mr
            Subjective+Objective parameters(all): & 0.23 & 0.07 & 0.24 & 0.11\\
            Distance, Duration, Velocity, Pace, MetricD, AHR, &  &  &  & \\
            MHR, Std, Skewness, Kurtosis, Metric1, Metric2 &  &  &  & \\
            Subjective + Objective parameters: & 0.39 & 0.27 & 0.32 & 0.13\\
            Distance, Duration, AHR, MHR &  &  &  & \\
            Subjective parameters (heart): AHR, MHR & 0.44 & 0.23 & 0.20 & 0.05 \\
            Subjective parameters (acc): Std, Skewness, Kurtosis & 0.59 & 0.67 & 0.99 & 0.99\\
            Metric1, Metric2 & & & & \\
            Subjective parameters (acc): Std, Skewness, Kurtosis & 0.63 & 0.56 & 1.07 & 1.16\\
            \br
        \end{tabular}
    \end{center}
\end{table}

\begin{figure}
\centering
        \includegraphics[height=7cm]{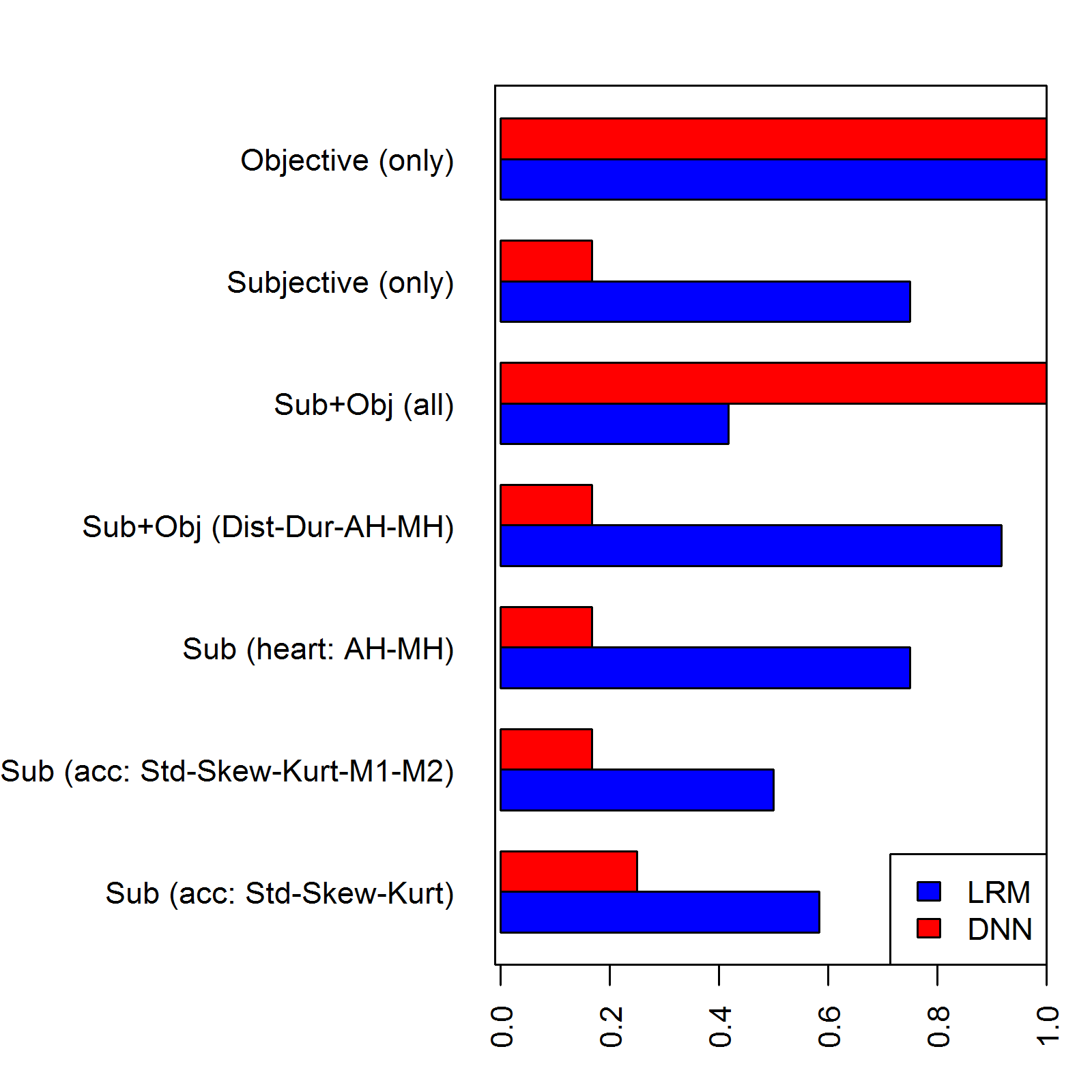}
        \caption{Accuracy of predictions made by LRM and DNN approaches for the various sets of parameters.}
        \label{Fig10_Accuracies}
\end{figure}

To compare the results obtained by deep learning neural network (DNN) with the standard approaches, the linear regression model (LRM) was used and the results of predictions made by LMR and DNN were compared by the confusion matrix and calculation of  accuracies (Fig.~\ref{Fig10_Accuracies}).

\begin{figure}
\centering
        \includegraphics[height=7cm]{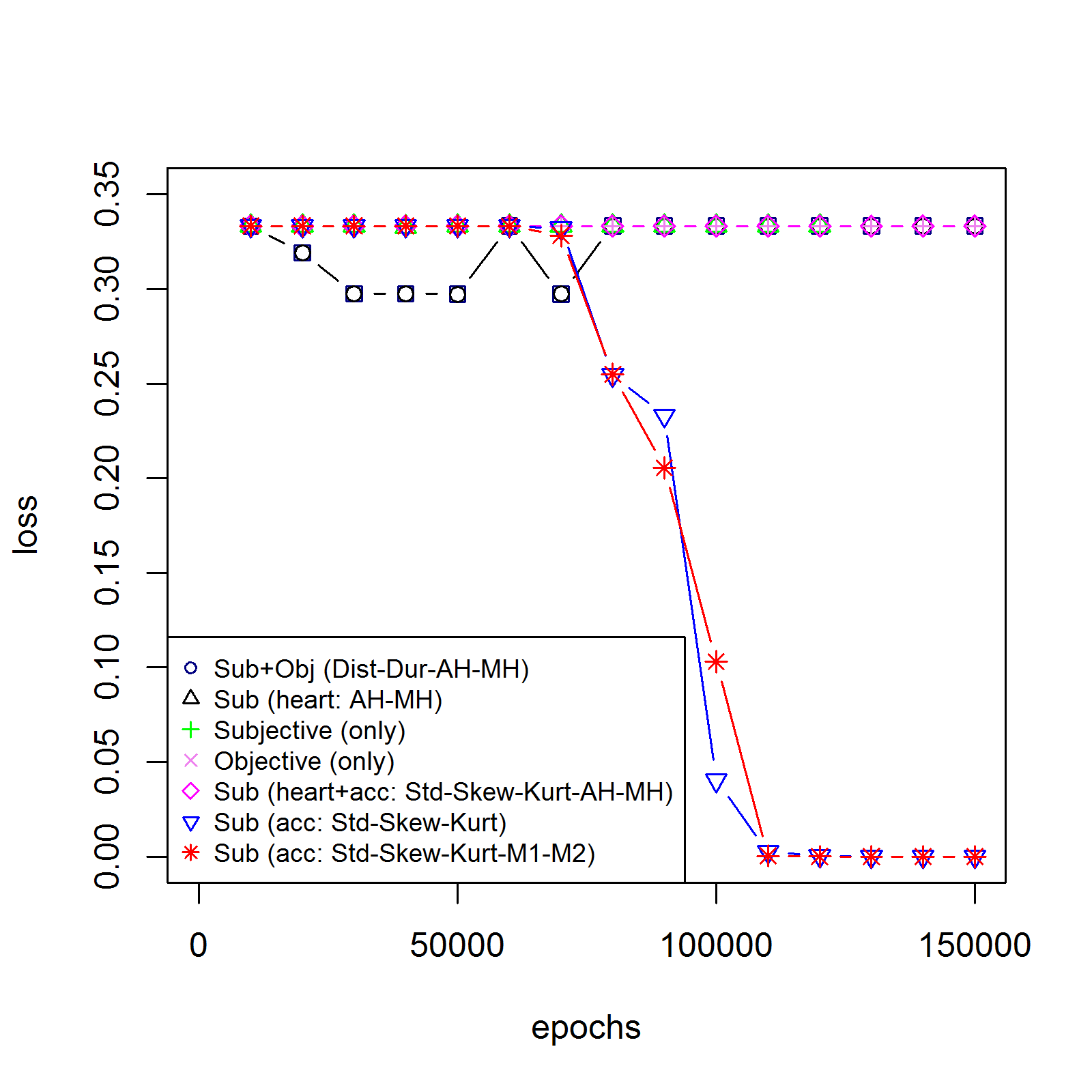}a) \includegraphics[height=7cm]{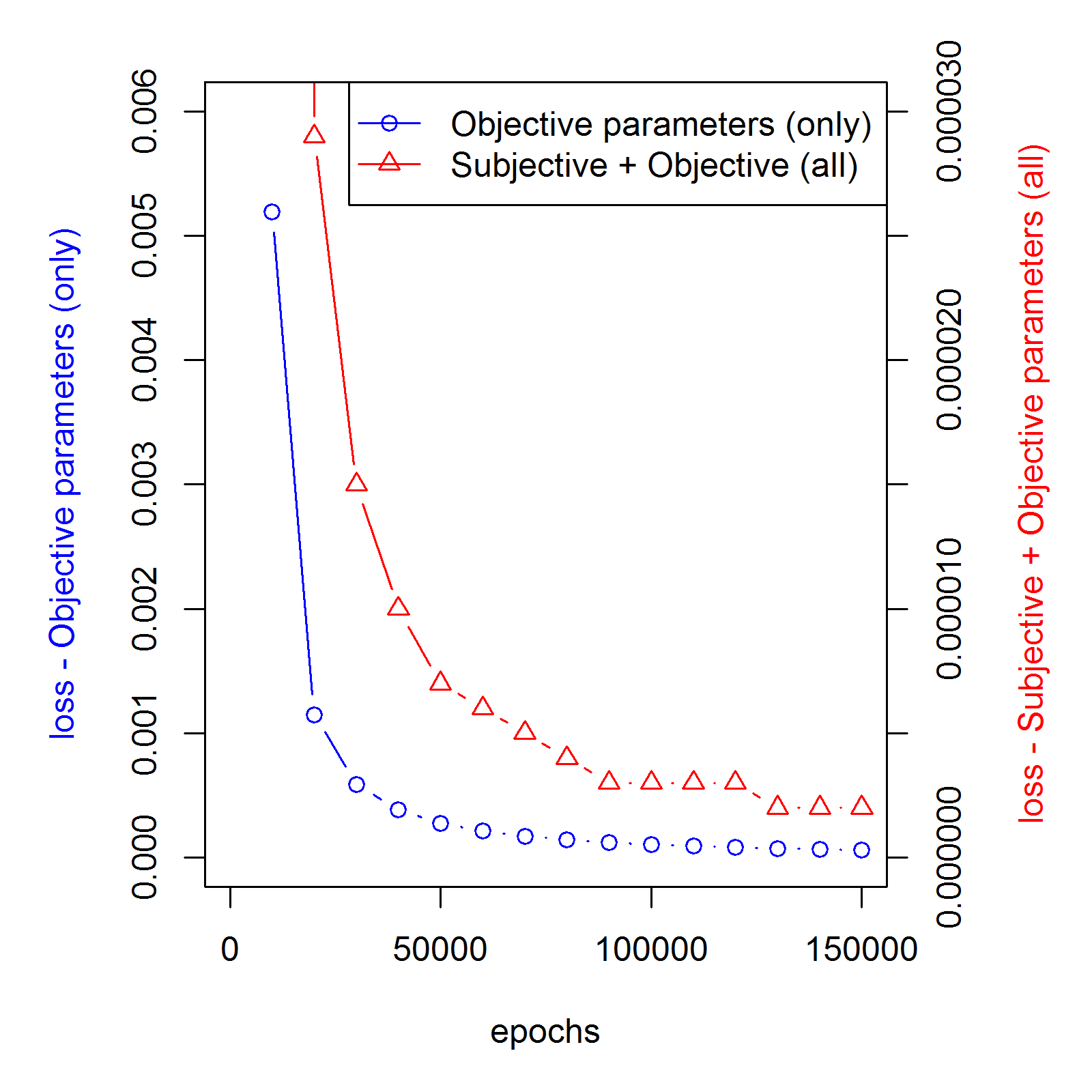}b)
        \caption{Loss during DNN training for various combination of parameters.}
        \label{Fig11_losses}
\end{figure}

Finally, analysis of losses was performed for both models with the various sets of parameters (Fig.~\ref{Fig11_losses}). Despite the tendency to learn from the training data, the loss is very high for most combinations of parameters, and the abrupt decrease of the loss for two of these combinations (blue and red colors in Fig.~\ref{Fig11_losses}a) is just illustration of over-training, but not the mark of the very reliable model. The more reliable results by DNN can be obtained for the bigger number of parameters like it was demonstrated in Fig.~\ref{Fig11_losses}b. They correspond to the highest accuracy values obtained by DNN (the first and third bars in Fig.~\ref{Fig10_Accuracies}).

\section{Discussions}
The obtained results had shown that some essential features of general activity can be distinguished and classified by the proper statistical analysis developed by us, namely: a) types of activity, i.e. from passive (sitting, writing, browsing, etc.) to active (walking, house keeping, etc.), and to very active (physical training, running, biking, etc.); b) quantitative estimation of intensity of activity; c) frequency and cumulative distribution of activities with time; d) energy spectrum of activities.

The tests with HB/HR monitor had shown that usage of HB/HR monitor allows us to find the following additional aspects: to monitor differences in heart activity (at exercise, recovery, rest) and distinguish low/high load patterns in heart recovery not only by direct comparison of the absolute values (heart rate and heartbeat), but by statistical analysis their distributions from the moments diagram. The results obtained by HB/HR monitoring could be especially valuable in combination with data on the well-known heart rate variability dependence on the physical load~\cite{gonzalez2017hrv}. 

These results allow us to make the previous conclusions that the parameters of the preprocessed data can provide the more targeted and sensitive characteristics (in comparison to raw data) of the response to the physical load, which was proved to be fruitful for other data mining purposes also~\cite{gordienko2015synergy}. Application of machine learning methods to train and recognize intensity of the aforementioned nonverbal communication allowed us to analyze the physical load on the persons under various objective parameters (for example, like distance, duration, velocity, etc. of walking, running, skiing) by these multimodal channels (acceleration, heart activity). But despite the current hype in deep learning domain, at the moment the results demonstrated here are not strong enough to be considered seriously for applications. The big potential of improvement is unavoidably related to availability of the much bigger dataset than the dataset used in this work. Also these results open several questions as to the possible ways for generation and estimation of the physical load and related fatigue in addition to other available work in this domain, especially on the basis of smartphones and other gadgets~\cite{lemoyn2017smartphone}. 

\section{Conclusions}
The main achievement is the multimodal data measured can be used as a training dataset for measuring and recognizing the intensity and physical load on the person by means of the machine learning approaches. The estimation of the mental load is the open question yet, because the 1-channel BCI device (MindWave Mobile by NeuroSky) is not precise and statistically reliable for the solid conclusions, but our previous analysis~\cite{stirenko2017user} shown that usage of the more powerful BCI devices (like multichannel OpenBCI~\cite{OpenBCI}) and EMG devices (like EMG-sensors by MyoWare~\cite{Myoware}) can be very promising in this context. The previous results demonstrate that usage of multimodal data sources (like wearable accelerometer, heart monitor, muscle movements monitor, brain-computer interface) along with machine learning approaches can provide the deeper understanding of the usefulness and effectiveness of this approach. In addition, one of the more interesting ways of further research would be creating algorithms for mapping behavioral patterns to ``action semantics'' with complex categories or structures like ``leisure activity after home or office or sport activity some time ago''. Nevertheless, the presented ideas can be applied for the further investigation of the physical load and related fatigue for the real use cases. It is especially important to increase the range of ways of understanding the rich set of other activities like eye blinking, facial gestures, body gestures, and body movements. In addition to the numerous other approaches~\cite{banaee2013data,lopez-nava2016wearable}, this approach open the new opportunities to exploit multimodal communication channels in different and personalized way, where each behavioral pattern can be trained by machine learning  and recognized for the purposes of users, for example, to help in health care for elders or people with disabilities.

\section{Acknowledgement}
The work was partially supported by Ukraine-France Collaboration Project (Programme PHC DNIPRO) (http://www.campusfrance.org/fr/dnipro) and Twinning Grant by EU IncoNetEaP project (http://www.inco-eap.net/).


\section*{References}
\begin{thebibliography}{00}

\bibitem{craighead2004concise} Craighead W and Nemeroff C 2004 {\it The concise Corsini encyclopedia of psychology and behavioral science} (John Wiley \& Sons)
\bibitem{Chernbumroong2013Elderly} Chernbumroong S, Cang S, Atkins A and Yu H 2013 Elderly activities recognition and classification for applications in assisted living {\it Expert Systems with Applications} {\bf 40} (5) 1662-1674
\bibitem{banaee2013data}Banaee H, Ahmed M and Loutfi A 2013 Data mining for wearable sensors in health monitoring systems: a review of recent trends and challenges {\it Sensors} {\bf 12} 17472-17500
\bibitem{lopez-nava2016wearable} López-Nava I and Muñoz-Meléndez A 2016 Wearable inertial sensors for human motion analysis: A review {\it IEEE Sensors Journal} {\bf 16} (22) 7821-7834
\bibitem{vinciarelli2015open} Vinciarelli A et al 2015 Open challenges in modelling, analysis and synthesis of human behaviour in human–human and human–machine interactions {\it Cognitive Computation} {\bf 7} (4) 397-413
\bibitem{meng2015review} Meng Y and Kim H 2012 A review of accelerometer-based physical activity measurement {\it Proc. Int. Conf. on IT Convergence and Security}  (Springer, Dordrecht) pp 223--237
\bibitem{lemoyne2010implementation} LeMoyne R, Mastroianni T, Cozza M, Coroian C and Grundfest W 2010 Implementation of an iPhone as a wireless accelerometer for quantifying gait characteristics {\it Annual Int. Conf. IEEE Engineering in Medicine and Biology Society} pp 3847-3851
\bibitem{gordienko2017augmented} Gordienko Yu et al 2017 Augmented Coaching Ecosystem for Non-obtrusive Adaptive Personalized Elderly Care on the Basis of Cloud-Fog-Dew Computing Paradigm {\it Proc. 40th Int. Convention on Information and Communication Technology, Electronics and Microelectronics} (Opatija, Croatia) pp 387-392 {\it Preprint arXiv}:1704.04988
\bibitem{stirenko2017user} Stirenko S et al 2017 User-driven Intelligent Interface on the Basis of Multimodal Augmented Reality and Brain-Computer Interaction for People with Functional Disabilities {\it Preprint arXiv}:1704.05915
\bibitem{weibull1951statistical} Weibull W 1951 Wide applicability {\it Journal of applied mechanics} {\bf 103} (730) 293-297
\bibitem{LeMoyne2015ml}LeMoyne R, Kerr W and Mastroianni T 2015 Implementation of machine learning with an iPod application mounted to a cane for classifying assistive device usage {\it Journal of Medical Imaging and Health Informatics} {\bf 5} (7) 1404-1408
\bibitem{lemoyne2018roleML} LeMoyne R and Mastroianni T 2018 Role of Machine Learning for Gait and Reflex Response Classification. {\it Wearable and Wireless Systems for Healthcare I. Smart Sensors, Measurement and Instrumentation} {\bf 27} (Springer, Singapore)
\bibitem{OpenBCI} OpenBCI official page (http://openbci./com); Cohen R 2014 New Open Source Platform Allows Anyone To Hack Brain Waves {\it Forbes} 2014-03-01 (Forbes Media)
\bibitem{Myoware} MyoWare Muscle Sensor, SparkFun (https://www.sparkfun.com/products/13723)
\bibitem{Hexiwear} NXP Accelerates Smart Wearable Product Development 2016 {\it Business Wire} 2016-02-23 (https://goo.gl/h3SF6Q)
\bibitem{gordienko2014change} Gordienko Y 2014 Change of scaling and appearance of scale-free size distribution in aggregation kinetics by additive rules {\it Physica A: Statistical Mechanics and its Applications} {\bf 412} 1-18
\bibitem{gordienko2012generalized} Gordienko Y 2012 Generalized model of migration-driven aggregate growth--asymptotic distributions, power laws and apparent fractality {\it International Journal of Modern Physics B} {\bf 26} (01) 1250010
\bibitem{gordienko2011molecular} Gordienko Y 2011 Molecular dynamics simulation of defect substructure evolution and mechanisms of plastic deformation in aluminium nanocrystals {\it Metallofizika i Noveishie Tekhnologii} {\bf 33} (9) 1217-1247
\bibitem{gonzalez2017hrv} Gonzalez K, Sasangohar F, Mehta R, Lawley M and Erraguntla M 2017 Measuring Fatigue through Heart Rate Variability and Activity Recognition: A Scoping Literature Review of Machine Learning Techniques {\it Proc. Human Factors and Ergonomics Society Annual Meeting} {\bf 61} (1) pp 1748-1752
\bibitem{gordienko2015synergy} Gordienko N, Lodygensky O, Fedak G, Gordienko Y 2015 Synergy of volunteer measurements and volunteer computing for effective data collecting, processing, simulating and analyzing on a worldwide scale {\it Proc. 38th Int. Convention on Information and Communication Technology, Electronics and Microelectronics} pp 193-198 {\it Preprint arXiv}:1504.00806
\bibitem{lemoyn2017smartphone} LeMoyne R and Mastroianni T 2017 Smartphone and portable media device: a novel pathway toward the diagnostic characterization of human movement {\it Smartphones from an Applied Research Perspective} ed N Mohamudally (InTech) DOI: 10.5772/intechopen.69961
\bibitem{kochura2017comparative} Kochura Y, Stirenko S,  Rojbi A, Alienin O, Novotarskiy M and Gordienko Y 2017 Comparative Analysis of Open Source Frameworks for Machine Learning with Use Case in Single-Threaded and Multi-Threaded Modes {\it XIIth Int. Scientific and Technical Conf. on Computer Sciences and Information Technologies} (Lviv, Ukraine) {\it Preprint arXiv}:1706.02248
\bibitem{kochura2017comparativeperformance} Kochura Y, Stirenko S and Gordienko Y 2017 Comparative Performance Analysis of Neural Networks Architectures on H2O Platform for Various Activation Functions {\it 2017 IEEE Int. Young Scientists Forum on Applied Physics and Engineering} (Lviv, Ukraine) {\it Preprint arXiv}:1707.04940
\bibitem{kochura2017ysf} Kochura Y, Stirenko S and Gordienko Y 2017 Performance Analysis of Open Source Machine Learning Frameworks for Various Parameters in Single-Threaded and Multi-Threaded Modes {\it Advances in Intelligent Systems and Computing} (Springer) {\it Preprint arXiv}:1707.08670

\end{thebibliography}
\end{document}